# Tunable superconductivity at the oxide-insulator/KTaO$_3$ interface and its origin

Changjiang Liu[1*], Xianjing Zhou[2], Deshun Hong[1], Brandon Fisher[2], Hong Zheng[1], John Pearson[1], Dafei Jin[2], Michael R Norman[1*], Anand Bhattacharya[1*]

[1]Materials Science Division, Argonne National Laboratory, Lemont, IL 60439, USA.
[2]Center for Nanoscale Materials, Argonne National Laboratory, Lemont, IL 60439, USA.

Email: changjia@buffalo.edu; norman@anl.gov; anand@anl.gov

Superconductivity forms out of the condensation of Cooper pairs of electrons[1]. The mechanism by which Cooper pairs are created in non-conventional superconductors is often elusive[2-5] because experimental signatures that connect a specific pairing mechanism to the properties of superconducting state are rare[6-8]. The recently discovered superconducting oxide-insulator/KTaO$_3$ interface[9] may offer clues about its origins. Here we observe distinct dependences of the superconducting transition temperature $T_c$ on carrier density $n_{2D}$ for electron gases formed at KTaO$_3$ (111), (001) and (110) interfaces. For the KTaO$_3$ (111) interface, a remarkable linear dependence of $T_c$ on $n_{2D}$ is observed over a range of nearly one order of magnitude. Further, our study of the dependence of superconductivity on gate electric fields reveals the role of the interface in mediating superconductivity, which also allows for a reversible electric switching of superconductivity at $T = 2$ K. We found that the extreme sensitivity of superconductivity to crystallographic orientation can be explained by Cooper pairing via inter-orbital interactions induced by the inversion-breaking transverse optical (TO1) phonons and quantum confinement. This mechanism is also consistent with the dependence of $T_c$ on $n_{2D}$ at the KTaO$_3$ (111) interface. Our study sheds light on the pairing mechanism in other superconducting quantum-paraelectrics.

**Introduction**

Ever since the advent of superconductors, investigations of the mechanism of superconductivity have been at the forefront of condensed matter physics. Recent years have seen the discovery of superconductivity in a plethora of engineered material systems, such as[10,11] LaAlO$_3$/SrTiO$_3$, FeSe/SrTiO$_3$ and twisted bilayer[12] and trilayer[13,14] graphene. These interfacial superconductors are of great interest due to the unconventional nature of their superconductivity, offering new routes towards pairing of electrons, as well as the tunability of their properties using electric field-effect gating[15]. For example, the presence of spin-orbit coupling and broken inversion symmetry at interfaces could enable new channels for Cooper pairing[16], and for realizing a pairing state[16,17] that can host zero-modes with non-Abelian statistics[18,19]. Insights into the pairing mechanisms in these material systems are thus of great interest, and an area of intense experimental and theoretical research.

In this work, we investigate the origins of superconductivity at oxide-insulator/KTaO$_3$ interfaces through chemical doping and electric field gating. We establish the doping dependence of $T_c$ for KTaO$_3$ (KTO) interfaces of different crystallographic orientations. A remarkable proportionality where $T_c \propto n_{2D}$ over nearly an order of magnitude is observed at the EuO/KTO (111) interface. In contrast, no superconductivity is observed at the (001) interface of KTO down to 25 mK for all $n_{2D}$, while the KTO (110) interface superconducts at a $T_c$ as high as 1K, intermediate between the (111) and (001) cases. Using electric field-effect gating, we tune both $n_{2D}$, as well as the confinement of electrons to the interface. We find that $T_c$ increases for both. We rule out the possibility that $T_c$ is controlled by the phase stiffness temperature[20]. We then discuss possible scenarios for pairing and find that inter-orbital interactions mediated by the soft transverse optical phonon reproduces key findings in our measurements, particularly the extreme sensitivity of $T_c$ to the crystallographic orientation of the interface, and the consistency with the linear dependence of $T_c$ with $n_{2D}$.

**Results**

**Distinct $n_{2D}$ dependences of $T_c$ at KTO (111), (001) and (110) interfaces**

Two-dimensional electron gases (2DEGs) at KTO interfaces are obtained by growing an oxide overlayer EuO on (111), (001) and (110) KTO surfaces using molecular beam epitaxy (MBE). The charge carriers at the EuO/KTO interface originate from chemical doping of oxygen vacancies and/or Eu substitution for K near the interface with KTO.[9] The carrier density $n_{2D}$ in these samples is determined using Hall measurements at 10 K. $n_{2D}$ of the as-grown samples can be tuned by nearly a factor of 10 by varying the growth conditions (see supplemental materials). We find that the donor state of charge carriers in these samples evolves gradually from shallow to deeper energy as $n_{2D}$ increases, the details of which are presented in the supplemental materials and Figure S1.

The EuO/KTO(111) samples are labelled as KTO_#, with # being in the order of increasing $n_{2D}$. Figure 1a shows the temperature dependence of the ratio of the sheet resistance $R_s/R_N$ ($R_N$ is the normal-state resistance at 4.5 K) for KTO (111) samples with different $n_{2D}$. The superconducting transition defined by $R_s/R_N$ going to zero occurs at progressively higher temperatures with increasing $n_{2D}$. Figure 1b shows the measurement of $R_s(T)$ on EuO/KTO (001) and (110) samples. As reported before[9], no superconducting transition can be measured for the (001) samples down to the lowest temperatures of ~ 25 mK. On EuO/KTO(110) samples, we observe that superconductivity emerges below about 1 K, which is similar to that reported for the LaAlO$_3$/KTO (110) interface[21]. The transition temperature $T_c$ for the KTO (110) samples also increases with $n_{2D}$, but gradually saturates.

When $T_c$ (determined at 20% of $R_N$) of all the KTO (111) samples are plotted against $n_{2D}$, as shown in Fig. 1c, we observe a largely linear dependence of $T_c$ on $n_{2D}$. The solid line is a fit using $T_c = a \cdot n_{2D} + b$, where the coefficient a ≈ 0.2 × 10$^{-13}$ K cm$^2$, and the y-axis intercept b ≈ - 0.1 K. We note that the fitting line extrapolates to near the origin of the axes, which implies that the undoped KTO (111) interface is near a doping induced critical point separating a band insulator and a superconducting phase. Within the entire doping range, $T_c$ versus $n_{2D}$ does not show a dome, suggesting that $T_c$ may be further enhanced by increasing the doping level. In Fig. 1d we plot $T_c$ versus $n_{2D}$ for KTO (001) and (110) interfaces, which show entirely different behavior than that of the KTO (111) interface: $T_c$ cannot be measured, down to 25 mK, for all $n_{2D}$ for the KTO (001) interface; while $T_c$ for the KTO (110) interface increases with $n_{2D}$ but saturates at a value about half that of the (111) case. We have also measured the Ginzburg-Landau coherence length ($\xi_{GL}$) in the KTO (111) samples with different $n_{2D}$, and found that $\xi_{GL}$ satisfies a scaling relation between $T_c$ and Hall mobility. The details are presented in the supplemental materials and Figure S2.

**Electrostatic tuning of superconductivity in a low $n_{2D}$ sample**

To gain further insight into the mechanism of the superconductivity, we have performed electrostatic gating measurements on the EuO/KTO (111) samples, in both the low and high $n_{2D}$ limits. It is expected that electric fields can tune a relatively large fraction of charge density in the low $n_{2D}$ sample in a back-gate geometry, given the large dielectric constant (~ 4500) of KTO at low temperatures[22-24]. Figure 2a shows a schematic of our measurement. A 100-nm thick Pt film deposited on the bottom side of KTO is used as the positive electrode, while the 2DEG at the EuO/KTO interface is grounded.

Figure 2b shows the measurements of $R_s(T)$ for different gate voltages $V_G$ on KTO_1, which has the lowest carrier density $n_{2D} = 1.25 \times 10^{13}$ cm$^{-2}$. For this sample, $R_N$ increases by over 3 times when $V_G$ is varied from 30 V to – 20 V. Correspondingly, $T_c$ becomes lower. We have measured $n_{2D}$, which increases with $V_G$ monotonically, as shown in the inset of Fig. 2c. By using a parallel-plate capacitor approximation and the thickness of KTO substrate of 0.5 mm, the slope of $n_{2D}$ vs. $V_G$ at $V_G = 0$ V yields a dielectric constant for KTO of ~ 5700, which is close to the known value for bulk KTO. When $T_c$ is plotted against $n_{2D}$ as tuned by $V_G$ (Fig. 2c), we observe a linear dependence of $T_c$ on $n_{2D}$, i.e., $T_c = 0.26 \times 10^{-13}$ (K cm$^2$)·$n_{2D}$ – 0.1 K. Crucially, the coefficient for $n_{2D}$ here is close to that found in the $n_{2D}$ dependence of $T_c$ obtained through chemical doping shown in Fig. 1c. The consistency between the electrostatic gating and chemical doping measurements highlights the predominant role of the carrier density, regardless of its origin, in tuning $T_c$ at the KTO interface. Figure 2d shows that the critical current $I_c$ also increases monotonically with $V_G$. We note that in recent ionic liquid gating measurements[25] on KTO, $T_c$ also changes monotonically with $V_G$, though the inferred $n_{2D}$ does not.

## Enhanced $T_{BCS}$ and a dome in $T_{BKT}$ by electrostatic gating

Next, we consider the effects of field-effect gating at higher carrier densities. Figure 3a shows $R_s(T)$ measured on KTO_9 with $n_{2D} = 1.04 \times 10^{14}$ cm$^{-2}$, for different $V_G$. As $V_G$ varies from 200 V to – 200 V, $R_N$ increases by about a factor of 2. At the same time, the superconducting transition occurs at higher temperatures, which is different than that seen on the low $n_{2D}$ sample. Utilizing the response of superconductivity to $V_G$, we show in Fig. 3b that the superconducting state with zero resistance can be switched on or off reversibly when $V_G$ is varied between – 75 V and 200 V at $T = 2.01$ K. We note that a superconducting state is maintained at low temperature over the entire range of $V_G$.

We found that the tunability of superconductivity here is due to an enhancement of the mean-field transition temperature $T_{BCS}$ with negative $V_G$. The transport data shown in Fig. 3a can be precisely interpreted by the Halperin-Nelson (HN) formula[26], which describes the increase in resistance with temperature due to the Berezinskii-Kosterlitz-Thouless (BKT) transition arising from superconducting phase fluctuations in 2D, as well as amplitude fluctuations of Cooper pairs at still higher temperatures as proposed by Aslamasov and Larkin (AL). The solid curve in Fig. 3c (and all those in Fig. 3a) is a fit to the data using $R_s(T) = R_N(1 + 4A^{-2}\sinh^2(b(T - T_{BKT})^{-1/2}T_{BKT}^{1/2}))^{-1}$, where A and b are fitting parameters that depend on the energy of the vortex core and the phase stiffness[27]. In the HN expression, $\Delta\sigma/\sigma_N = 4A^{-2}\sinh^2(b(T - T_{BKT})^{-1/2}T_{BKT}^{1/2})$ is a measure of the enhanced conductivity by superconducting fluctuations. Figure 3d shows a contour plot for $\Delta\sigma/\sigma_N$ as a function of $V_G$ and temperature $T$, which is obtained from the HN analysis for measurements over the full range of $V_G$. As $V_G$ goes negative, $\Delta\sigma/\sigma_N$ shows monotonic enhancement for $T \geq 2.15$ K, while at lower temperatures a local maximum in $\Delta\sigma/\sigma_N$ is seen.

From the HN analysis, we obtain the $T_{BKT}$, which is found to be close to $T_{c0}$, where $R_s$ is zero. The mean-field $T_{BCS}$ is obtained from the inflection point[28] of the HN fit (maximum of its first derivative, near 20% of $R_N$), and has been confirmed by AL fits (Figure S3). Figure 3e shows how $T_{BCS}$ and $T_{BKT}$ evolve as a function of $V_G$. $T_{BCS}$ is monotonically enhanced as $V_G$ goes to negative values, while $T_{BKT}$ shows a local maximum at $V_G \sim -75$ V. We also measured the voltage-current (V-I) characteristic near $T_{c0}$ to determine $T_{BKT}$ independently. As shown in Fig. 3f, $T_{BKT}$ obtained through V-I measurements at $V_G = 0$ V is $\sim 2.01$ K, very close to the results obtained by the HN fit. See also Figure S4 for V-I measurements at different $V_G$.

We find that the non-monotonic dependence of $T_{BKT}$ on $V_G$ is caused by the enhancement of $T_{BCS}$ together with the increase of $R_N$ as $V_G$ decreases. The ratio of $T_{BKT}/T_{BCS}$ from our data, as shown in Fig. 3g, follows closely the prediction for a 2D superconductor[29], i.e., $T_{BKT}/T_{BCS} = (1 + 0.173R_N/R_c)^{-1}$, where $R_c = \hbar/e^2$ with $\hbar$ being the reduced Planck constant. The increase of $R_N$ by negative $V_G$ is mainly due to charge carriers being pushed closer to the EuO/KTO interface, where more disorder is present[30]. As shown in Fig. 3h, the Hall mobility decreases from about 37 cm$^2$ V$^{-1}$ m$^{-1}$ to 17 cm$^2$ V$^{-1}$ m$^{-1}$ as $V_G$ varies from 200 V to $-200$ V, while the variation in carrier density from Hall measurements is only $\sim 5\%$. These observations show that, in addition to $n_{2D}$ as found earlier, the proximity of the carriers to the interface also increases $T_c$ ($T_{BCS}$), suggesting the interface itself is the origin of pairing. We also observe that the critical field is enhanced at negative $V_G$, which is found to be mainly due to the decrease in mobility (Figure S5).

**Pairing mechanism at the KTO interfaces**

We first discuss the marked proportionality of $T_c$ with $n_{2D}$ observed for the KTO (111) interface. One possible explanation for this phenomenon is that $T_c$ is limited by the phase stiffness of the superconducting state, where $T_c \propto E_J$ ($T_{BKT}$) with $E_J = \hbar^2 n_s(T)/4m^*$, $n_s$ the 2D superfluid density, $m^*$ the effective mass of the electrons, and $n_s(0\text{ K}) \propto n_{2D}$. This would give a linear relation between $T_c$ and $n_{2D}$. However, our analysis shown in Fig. 3c indicates that the temperature range over which BKT physics is dominant is small. That is, over most of the temperature range, the resistance change is due to amplitude fluctuations of Cooper pairs. In this context, we note that $T_{BCS}$ is only slightly larger than $T_{BKT}$.

The above discussion leads to the second possibility for the observed proportionality between $T_c$ and $n_{2D}$, which is that the pairing interaction is non-BCS like with a superconducting gap $\Delta \propto E_F$ (Fermi energy) instead of $\hbar\omega_D$ (Debye energy). Such a situation would occur in the anti-adiabatic limit[31] with $\hbar\omega_D > E_F$. Since in 2D, $n_{2D}$ scales with $E_F$, one would obtain $T_c \propto \Delta \propto n_{2D}$. The estimated $E_F$ in our samples lies in the range of 10 - 80 meV, so pairing via the highest-energy longitudinal optical mode (LO4) with energy ~ 100 meV would be required. There are several potential issues with such a scenario. First, whether $\hbar\omega_D$ can by replaced by $E_F$ in the BCS expression for $T_c$ is controversial[5]. Second, it would require that the BCS coupling constant be independent of $n_{2D}$.

This leads us to consider pairing via the soft transverse optical (TO1) phonon mode[32], the phonon responsible for the nearly ferroelectric behavior of KTO. We note that the standard gradient coupling of the electrons to TO modes vanishes as the momentum transfer $q$ goes to 0, meaning that only pairs of phonons can couple to the electrons[33]. However, as pointed out in recent work[34,35], inversion breaking from the TO1 mode leads to linear coupling to electrons for inter-orbital interactions among the three $t_{2g}$ orbitals: $d_{xy}$, $d_{yz}$, and $d_{zx}$, which is illustrated in Fig. 4a. The displacements of Ta and O sites in the vertical direction, indicated by the black arrows, break inversion symmetry with respect to the horizontal plane joining the center of the $d$ orbitals. As a result, the inter-orbital hopping of electrons along a Ta-O-Ta bond is no longer zero.

We show in detail in the supplemental materials that: (*i*) The TO1 mode energy ($\omega_{TO1}$) increases as a function of $n_{2D}$ due to the electric field screening provided by the charge carriers. (*ii*) The BCS coupling constant λ (arising from the Rashba-like splitting of the energy bands with wavevector $k$) also increases with $n_{2D}$. We take $T_c = 1.14\ \omega_{TO1}(q = 2k_F, n_{2D})\ \exp[-(1+\lambda)/\lambda]$. Adjusting $\lambda = c\ n_{2D}\ <1/\omega_{TO1}^2>$ by a free parameter $c$ to scale the value of $T_c$ to equal the experimental value of ~ 2 K at $n_{2D} = 1 \times 10^{14}/cm^2$, we obtain a nearly linear dependence of $T_c$ on $n_{2D}$ for the KTO (111) interface, which is shown in Fig. 4b. Here, $<>$ is a Fermi surface average (see supplemental materials) and we assume that $\omega_{TO1}^2(q=0) = c_1 + c_2\ n_{2D}$ as observed in SrTiO$_3$, with $c_1$ and $c_2$ being constants.

This scenario also predicts the absence of superconductivity for the KTO (001) interface, i.e., the superconductivity is highly sensitive to the crystallographic orientation. This is due to the nature of the inter-orbital electron-TO1 phonon coupling and the effects of quantum confinement for electrons at the KTO-oxide interface. The degeneracy of the $t_{2g}$ orbitals for the 2DEG is lifted by quantum confinement depending on the crystalline surface normal, which is illustrated in Figs. 4c - e for the three KTO surface orientations. This lifting of degeneracy reduces the number of $d$ orbitals participating in the inter-orbital hopping. Consequently, electrons and TO1 phonons are maximumly coupled for the KTO (111) surface where the three $t_{2g}$ orbitals are degenerate, while they are largely decoupled for the KTO (001) surface given the splitting between the $d_{xy}$ and $d_{xz/yz}$ orbitals. Shown in Fig. 4f are the calculated energy bands for the KTO (111) surface exhibiting large 'Rashba'-like splitting due to inter-orbital hopping, while the energy bands for the KTO (001) surface shown in Fig. 4g do not exhibit a noticeable splitting. Further, we find that the electron-TO1 phonon coupling strength for the KTO (110) surface is intermediate because the inversion breaking does not occur along the [001] crystal axis (Fig. 4e) in the (110) plane, reflecting the reduced orbital degeneracy of (110) relative to (111).

A more detailed analysis of pairing in KTO is highly non-trivial. It would also need to account for splitting of the TO1 mode due to both symmetry breaking and the electric field at the interface. As the inversion breaking is largest at the interface, this would be consistent with our observed increase of $T_{BCS}$ when the charge carriers are pushed closer to the interface by a negative $V_G$. A detailed analysis also needs to consider (1) the influence of both LO and TO modes on $T_c$, (2) the carrier distribution $n_{3D}(z)$ along the interface normal ($z$) together with the strong interface orientation dependent electronic structure[36-38], and (3) the influence of inter-orbital terms between the $t_{2g}$ and $e_g$ orbitals. We note that while the dependence of $T_c$ on crystal orientation is not observed[10,39,40] in SrTiO$_3$ based 2DEGs (barring a recent exception[41]), this may be due to a weaker lifting[42-44] of the degeneracy of the $3d$ orbitals (the confinement potential being weaker in STO than KTO). We also note that while the $t_{2g}$ orbitals in bulk $KTaO_3$ are degenerate, it is not known to superconduct. We believe this may be due to the relatively low density of states in doped bulk KTO samples (in 3D, the density of states is proportional to $E^{1/2}$ whereas in 2D it is a constant), as well as the possibility of polaron formation.

## Supplemental materials

**Material growth.** The growth steps include annealing the substrate at a temperature ~ 600 °C, and then exposing it to a flux of Eu atoms in the temperature range 375°C - 550°C, both carried out for pressures in the $10^{-10}$ Torr range. The annealing process promotes O vacancy formation near the surface, and subsequent exposure to Eu causes the formation of a thin layer of EuO$_x$, as the Eu scavenges O from the substrate[9]. Following this step, the EuO overlayer is grown under a O$_2$ partial pressure between 0.5 – 10 x $10^{-9}$ Torr with a final EuO thickness of ~ 25 nm. Our microscopy results[9] indicated the presence of O vacancies and Eu substitution on the K sites in KTO, which would both promote the formation of an interfacial electron gas. Upon varying the temperature at which the KTO substrate is exposed to Eu in vacuum, we were able to vary the doping levels in the 2DEG. We note that the EuO/KTO interface can show transport anisotropy or a 'stripe'-like phase in the (111) plane[9]. However, this does not influence the analysis of $T_c$, at which global superconductivity sets in as that is found to be the same for 'stripe' like samples at a given value of $n_{2D}$. We also find that in samples with 'stripe' like behavior the anisotropy is strongly reduced when patterned into Hall bars vs those that are measured using a van der Pauw geometry, details of which are currently being studied. The samples presented here show weak or no anisotropy.

**Evolution of donor states as a function of doping.** $n_{2D}$ in these samples is determined using Hall measurements at 10 K. Figure S1a shows the Hall resistance $R_H$ measured as a function of magnetic field for different KTO (111) samples. Notably, the field dependences of $R_H$ are all linear, with negative slopes, consistent with electron-like charge carriers. We found that the donor state of charge carriers evolves gradually from shallow to deeper energy as $n_{2D}$ increases. Figure S1b shows the temperature dependence of $n_{2D}$ in the range 300 K – 2 K. For samples with $n_{2D}(10\text{ K}) \lesssim 5 \times 10^{13}$ cm$^{-2}$, the measured $n_{2D}(T)$ increases upon cooling. This suggests that charge carriers in these samples mainly come from shallow donors whose energies merge with the conduction band due to the nearly 20-fold increase in the dielectric constant in KTO upon cooling. For samples with $n_{2D}(10\text{ K}) \gtrsim 5 \times 10^{13}$ cm$^{-2}$, $n_{2D}(T)$ decreases upon cooling. This suggests that at higher doping levels charge carriers originate from deeper donor states[45], to which some of them freeze out upon cooling. Figure S1b also shows that regardless of the doping level in the as-grown samples, $n_{2D}$ remains nearly constant for temperatures below 10 K for all samples.

**Scaling relation of coherence length.** We have measured the Ginzburg-Landau coherence length ($\xi_{GL}$) in the KTO (111) samples with different $n_{2D}$, by measuring the out-of-plane upper critical field $B_{c2}$ as a function of temperature. $B_{c2}$ shows a linear temperature dependence near $T_c$, from which we calculate $\xi_{GL}$ using $B_{c2}(T) = \Phi_0(1 - T/T_c)/[2\pi(\xi_{GL})^2]$, where $\Phi_0$ is the magnetic flux quantum. Figure S2a shows the data of $B_{c2}$ versus $T$ for five samples. The $B_{c2}$ varies greatly among these samples with different $n_{2D}$ and $T_c$. Figure S2b summarize the $\xi_{GL}$ for all the samples measured, which lies in the range from 10 to 57 nm. We found that $\xi_{GL}$ scales with $(\mu/T_c)^{1/2}$. As shown in Figure 2b, the ratio of $\xi_{GL}$ to $(\mu/T_c)^{1/2}$ remains constant for samples over the entire $n_{2D}$ range. We found that this relation also holds when the superconductivity is tuned by $V_G$. See Figure S4.

For a BCS superconductor in the dirty limit, it is expected that $\xi_{GL} \propto (l\xi_{BCS})^{1/2}$ with $\xi_{BCS}$ being the BCS coherence length[46]. Using $l \propto v_F\mu$ and $\xi_{BCS} \propto v_F/T_c$ with $v_F$ being the Fermi velocity, one gets $\xi_{GL} \propto v_F(\mu/T_c)^{1/2}$. For the KTO (111) interface, we observe that $\xi_{GL} \propto (\mu/T_c)^{1/2}$ instead. This would imply that $v_F$ is a constant, independent of $n_{2D}$. One way this could happen is if there is a linear dispersion of energy versus momentum like in graphene, though this conflicts with the known electronic structure of the KTO conduction bands that have a parabolic dispersion around the $\Gamma$ point.

**Calculation of $T_c$ versus $n_{2D}$ from TO1 mode pairing.**

**Evaluation of TO1 mode energy.** The BCS formula for pairing from a TO1 mode is taken to be $T_c = 1.14\ \omega_{TO1}(q = 2k_F, n_{2D}) \exp[-(1 + \lambda)/\lambda]$ where $\lambda$ is the BCS coupling constant (any Coulomb pseudopotential, $\mu^*$, has been ignored). We first discuss the assumptions behind this expression, and then how to evaluate the various quantities. The first assumption is that the relevant pairing scale is controlled by the momentum transfer $q$ of the TO1 mode for scattering around the Fermi surface, hence we set the cut-off energy to the maximum mode energy which is at $q = 2k_F$. Therefore, the $n_{2D}$ dependence of $T_c$ will come from any hardening of the TO1 mode along with the increase of $k_F$ with $n_{2D}$ since the TO1 mode has a substantial dispersion. The TO1 mode dispersion is taken from the low energy neutron scattering study of Farhi *et al.*[47] There, the phonon dispersion is fit with an expression due to Vaks, and in the approximation where the anisotropy of the dispersion is ignored involves solving a 2 by 2 matrix that couples the transverse acoustic mode to the transverse optic one. We do this using the parameters of Ref. [47] evaluated at their lowest temperature of measurement (10 K). To proceed further, we need to know the $n_{2D}$ dependence of the mode energy at $q = 0$, as well as that of $k_F$. For the former, we will assume that $\omega_{TO1}^2(q = 0, n_{2D}) = c_1 + c_2 n_{2D}$; this relation is found in STO where doping of the bulk with free carriers has been observed (KTO is resistant to doping in the bulk). The resulting mode hardening is consistent with field dependent Raman data by Fleury and Worlock[48]. $c_1$ is set by the bulk value of $\omega_{TO1}$ of 2.5 meV. We set $c_2$ by the estimated value of $\omega_{TO1}$ at $n_{2D} = 10^{14}$ cm$^{-2}$ of 5.6 meV. The latter has been obtained from Ueno et al.[49] that relates the field dependent dielectric function to $n_{2D}$, with the relative dielectric function expressed as $\varepsilon(E) = 4500/(1 + b\ E)$ with $b = 8 \times 10^{-7}$ ($E$ in V m$^{-1}$). The relevant "average" field $F$ is then determined as in Ref. [50] by assuming a triangular confining potential along $z$ (the normal to the interface), and integrating the dielectric constant with respect to field up to $F$. That is, $e \cdot n_{2D} = 2\int_0^F \varepsilon_0 \cdot \varepsilon(E) dE$, where $e$ is elementary charge and $\varepsilon_0$ is 8.85 pF m$^{-1}$. The mode energy is then given by the Lyddane-Sachs-Teller relation $\omega_{TO1}^2(q = 0, n_{2D}) = \omega_{TO1}^2(q = 0, n_{2D}=0)\ \varepsilon(0)/\varepsilon(F)$. The reason we did not use this formalism over the entire range of $n_{2D}$ is that the resulting $n_{2D}$ dependence of $\omega_{TO1}^2(q = 0)$ deviates from the above assumed linear behavior (Figure S6e), which in turn has a detrimental impact on the functional dependence of $T_c$ on $n_{2D}$ (Figure S6f). In the future, this could be looked at by measuring the actual TO1 mode energy as a function of $n_{2D}$ in KTO (realizing that the TO1 mode polarized normal to the interface should harden more with $n_{2D}$ than the TO1 mode orthogonal to this).

To obtain the variation of $k_F$ with $n_{2D}$, we assume the simple tight binding model of Ref. [51], where the electronic structure is that of a (111) bilayer, with the parameters determined in order to reproduce the ARPES data from the (111) KTO surface of Ref. [38]. This results in a near neighbor hopping energy $t$ of 1 eV, with the spin-orbit coupling $\xi = 0.265$ eV (in order to reproduce the quartet-doublet spin-orbit splitting of 0.4 eV at $\Gamma$). For $k_F$, we take that of the larger of the two Fermi surfaces (the quartet splits into two doublets upon dispersing away from $\Gamma$), and set $k_F$ to its value along $\Gamma$-K of the surface Brillouin zone. $n_{2D}$ is gotten by determining the occupied area of the two Fermi surfaces at $E_F$. The resulting dependence of $E_F$ and the TO1 mode energies with $n_{2D}$ is shown in Figure S6a. One can see the approximate linear dependence of $E_F$ with $n_{2D}$ (expected for a parabolic dispersion in 2D), with the deviation at larger $n_{2D}$ due to the approach to a van Hove singularity of the lower band of the quartet that occurs at the M point of the surface zone. As one can also see, the dependence of the mode energy at $q = 2k_F$ is not linear in $n_{2D}$. This is to be expected, since in the approximation where the mode energy at $q = 0$ is small compared to that at $q = 2k_F$, the mode energy should scale with $q$ and thus $k_F$, the latter going approximately as the square root of $n_{2D}$. Also note that over most of the $n_{2D}$ range, the mode energy at $q = 2k_F$ is much smaller than $E_F$, justifying the BCS (adiabatic) approximation for TO1 mode pairing.

**BCS coupling constant $\lambda$.** For TO1 mode pairing, $\lambda$ is given by[35]: $\lambda = c\, n_{2D} <1/\omega_{TO1}^2(q, n_{2D})>$ where $<\,>$ represents an average over the Fermi surface. As the mode dispersion is quadratic in $q$, the Fermi surface average is trivial, resulting in $\lambda = c \cdot n_{2D}/[\omega_{TO1}(q = 2k_F, n_{2D}) \cdot \omega_{TO1}(q = 0, n_{2D})]$. Here, $n_{2D}$ in the numerator comes from $k_F^2$ ($k_F^2 = 2\pi \cdot n_{2D}$). This factor arises from the electron-phonon vertex that is proportional to $k_F$ (that is, the effect of inversion breaking from the polar TO1 mode leads to a linear splitting of the energy bands with $k$). More will be described about this inversion breaking effect in the next section. The constant $c$ subsumes the proportionality coefficient of the vertex along with the density of states at the Fermi energy, $N_F$, that we assume is constant (equivalent to linearity of $E_F$ with $n_{2D}$). This is in contrast to the bulk[52] where $N_F$ goes as $n_{3D}^{1/3}$. The value of $c$ would require a detailed microscopic theory. In lieu of that, we set $c$ in order to give $T_c$ of ~ 2 K at $n_{2D} = 10^{14}$ cm$^{-2}$ (for this $n_{2D}$, $\lambda \sim 0.26$ and so $c = 2.85$ when $n_{2D}$ is in units of $10^{13}$ cm$^{-2}$ and $\omega_{TO1}$ in meV). The resulting $\lambda$ and $T_c$ versus $n_{2D}$ is plotted in Figure S6b and c, respectively. $T_c$ is then replotted in Figure S6d assuming that $\lambda$ is reduced by a factor of 2.

One can see a remarkable linear variation of $T_c$ with $n_{2D}$ in Figure S6c, despite the non-linearity with $n_{2D}$ of both the BCS cut-off and the BCS coupling constant. In Figure S6d, one sees a strong suppression of $T_c$, with values similar to those claimed by Ueno *et al.* for the 001 interface[49] (a further reduction in $\lambda$ would completely suppress $T_c$ as we observe). Since $\lambda$ is proportional to the square of the electron-phonon vertex, a mere reduction of the vertex by $2^{1/2}$ is sufficient to cause an enormous reduction in $T_c$. As such, we expect $T_c$ to be extremely sensitive to the interface orientation, which we address next.

**Dependence of $T_c$ on interface orientation from TO1 mode pairing.**

**Electron-phonon vertex.** To obtain a significant electron-phonon vertex for a TO1 mode, one requires inter-orbital terms (intra-orbital ones instead lead to a coupling quadratic in the ion displacement[33]). Therefore, in a TO1 mode picture, one requires orbital degeneracy in order to get a substantial $T_c$. This qualitatively explains the observed dependence of $T_c$ with interface orientation. For the (111) case (ignoring any trigonal distortion), the three $t_{2g}$ states are degenerate at Γ (being split into a lower quartet and an upper doublet by spin-orbit). We can contrast this with the (001) case, where confinement along [001] leads to the $d_{xy}$ state being pulled down substantially relative to the $d_{xz/yz}$ ones. This splitting acts to suppress the effect of inter-orbital terms, consistent with our lack of observation of superconductivity on (001) interfaces (in Ref. [49], a $T_c$ of less than 50 mK was observed). For (110), we expect something intermediate, because the effect of confinement in that case is to instead pull down the $d_{xz/yz}$ states relative to the $d_{xy}$ one (so, partial degeneracy).

To go beyond these qualitative observations, we need to consider the variation of the electronic structure with interface orientation, and we also need an estimate of the coupling of these electrons to the TO1 mode. To address the latter, we follow Ref. [35], where the TO1 mode is considered to be a Slater mode (known to be a good approximation for the TO1 eigenvector). As they show, there are two contributions. One is the standard gradient one which only involves the metal ion motion and disappears as $q$ goes to 0. The second is the inter-orbital terms that involves primarily the oxygen ion motion and does not vanish as $q$ goes to 0.

The inversion splitting of the bands near Γ due to the TO1 mode gives rise to an electron-phonon vertex[35] $g_{TO1} = 2t'c'dk_Fa$ where $t'$ is the derivative of the near-neighbor hopping with respect to the ion displacement, $c' = (\eta^{-1/2} + \eta^{1/2})$ where $\eta = 3M_O/M_{Ta}$, $d$ is the zero-point displacement of the ions from excitation of the Slater mode given by $d = (\hbar/2M_S\omega_{TO1})^{1/2}$, $M_S$ is the mass of the Slater mode ($3M_O+M_{Ta}$), and $a$ is the lattice constant (3.9884 Å for KTO). The existence of $t'$ allows coupling between one $t_{2g}$ orbital on a given Ta site with a different $t_{2g}$ orbital on a neighboring Ta site (which is not allowed if the Ta-O-Ta bond is linear). $t'$ was obtained in Ref. [35] from a frozen phonon calculation. Instead, we follow Ref. [34] which calculates the inter-orbital coupling due to displacement of the oxygen ion transverse to the Ta-O-Ta bond. This leads to a value of $t' = 2t_{pd}^2/(\Delta_{pd}a)$ where $t_{pd}$ is the $pd\pi$ hopping and $\Delta_{pd}$ the energy difference between the O 2p and Ta 5d orbitals, noting that the oxygen ion is at a distance $a/2$ relative to the Ta ion. $t_{pd}$ and $\Delta_{pd}$ can be obtained from Mattheiss' Slater-Koster fits[53] for KTO. Doing this we obtain the values listed in Table S1. $t_R$ is the inversion breaking hopping parameter that enters the electronic structure Hamiltonian[51], and the TO1 mode energy listed is that of the bulk insulator at $q = 0$. The reason we included STO is to compare our crude approximation of $t'$ to the more sophisticated one of Ref. [35]. Our value for STO is about 50% larger than theirs (our value for KTO being similar to theirs for STO). The reason the STO value is larger is because of the softer TO1 mode coupled with the larger $d$ (zero-point motion) due to the lighter Ti ion.

On the other hand, there are other inversion breaking terms that have been ignored here, and those couple the $t_{2g}$ electrons to the $e_g$ ones[54]. As those terms are proportional to the inversion breaking times the spin-orbit coupling divided by the $t_{2g}$-$e_g$ splitting, they are presumably much larger in KTO than STO given its 20 times larger spin-orbit coupling $\xi$. These terms also exhibit orbital differentiation. For instance, for displacements along [001], the $d_{xy}$ states do not couple to the $e_g$ states, and therefore similar considerations to those above apply for these $t_{2g}$-$e_g$ terms as well. These terms in turn could explain why $T_c$ in KTO is much higher than in STO. But their inclusion results in a much more involved theory than that presented here, so we leave this to future work. In this context, note that our dispersion plots are based on displacements along the interface normal. In the dynamic case relevant for the calculation of $T_c$, one or both TO1 modes (or neither) come into play depending on the specific orbitals and interface orientation involved.

Regardless, the purpose of this exercise was to give a rough estimate of the size of $t_R$ due to dynamic coupling of the electrons to the TO1 mode. Its value above (~ 32.5 meV) is extremely large, as commented on in Ref. [35], meaning that TO1 mode pairing is quite viable. In Fig. 4f, g and Figure S7a, we show the electronic structure in the bilayer approximation for the three interface orientations assuming this value of $t_R$.

**Splitting in the energy dispersion and interface orientation variation of $T_c$.** For KTO (111) (Fig. 4f), note the pronounced Rashba-like effect near $\Gamma$ (displaced parabola) for the lowest band of the spin-orbit quartet. For (001) (Fig. 4g), note the small effect of inversion breaking. For (110) (Figure S7a), note the small splitting along $\Gamma$-Z (similar to (001)) and the much larger splitting along $\Gamma$-M, along [1-10]. This (110) case leads to a large Fermi surface anisotropy, while the (001) case is isotropic and the (111) case only mildly anisotropic, since the Ta-O-Ta bonds are parallel to [001] but are not parallel to [1-10] in the (110) plane. These electron-TO1 mode couplings for different KTO surfaces are schematically illustrated in Fig. 4c-e of the main text using dark and light colors for the Ta-O-Ta bonds.

We now address the interface orientation variation of $T_c$ from the variation of the inversion breaking effect with orientation. Note that the $k_F^2$ factor in the numerator of the BCS coupling constant $\lambda$ comes from the square of the splitting of the bands due to inversion breaking, the latter going like $\alpha_R k_F$, with the notation $\alpha_R$ denoting that many (but not all) of these inter-orbital terms having a Rashba-like form. So, $\lambda$ scales as $\alpha_R^2$. $\alpha_R$ is equal to a constant times $t_R$. We denote this constant by $\delta$ (which has units of Angstroms). So, $\lambda$ scales as $\delta^2$. In Table S2 are values of $\delta$ gotten from the slope of the lowest energy band versus $t_R$ divided by $|k|$ ($k$ is taken to be near $\Gamma$), which we extract from that of the lowest band. An example is shown in Figure S7b.

The first thing to note from Table S2 is that (001) will have a negligible $\lambda$. In fact, we can get an analytic estimate of what the $\delta$ ratio is from Ref. [54]. The inversion breaking term for (111) is proportional to $3^{1/2}$ (ignoring coupling to $e_g$ states). The inversion breaking for (001) is proportional instead to $2\xi/\varepsilon$ where $\varepsilon$ is the splitting of the $d_{xy}$ state from $d_{xz/yz}$. Therefore, the ratio of $\delta$ (111 to 001) is $3^{1/2}\varepsilon/2\xi$. In our simple model, $\varepsilon$ is equal to $t$. For our values ($t = 1$ eV, $\xi = 0.265$ eV), we get a ratio of 3.27, which approximately matches the numerical value from Table S2 (3.78). Since $\lambda$ goes as the square of this ratio, one sees that $T_c$ on the (001) interface should be negligible.

For (110), it is difficult to give an estimate given the large anisotropy of the inversion breaking in the surface Brillouin zone coupled with the large anisotropy of the Fermi surface. Still the (110) $\delta$ values are much closer to the (111) value than the (001) value, so a $T_c$ for 110 half that for (111) is reasonable. The upshot is that the $T_c$ variation follows the breaking of orbital degeneracy, as shown by the last two columns of Table S2. Here, "orbitals" refer to the $t_{2g}$ orbitals present in the low energy manifold of states.

**Main Figures**

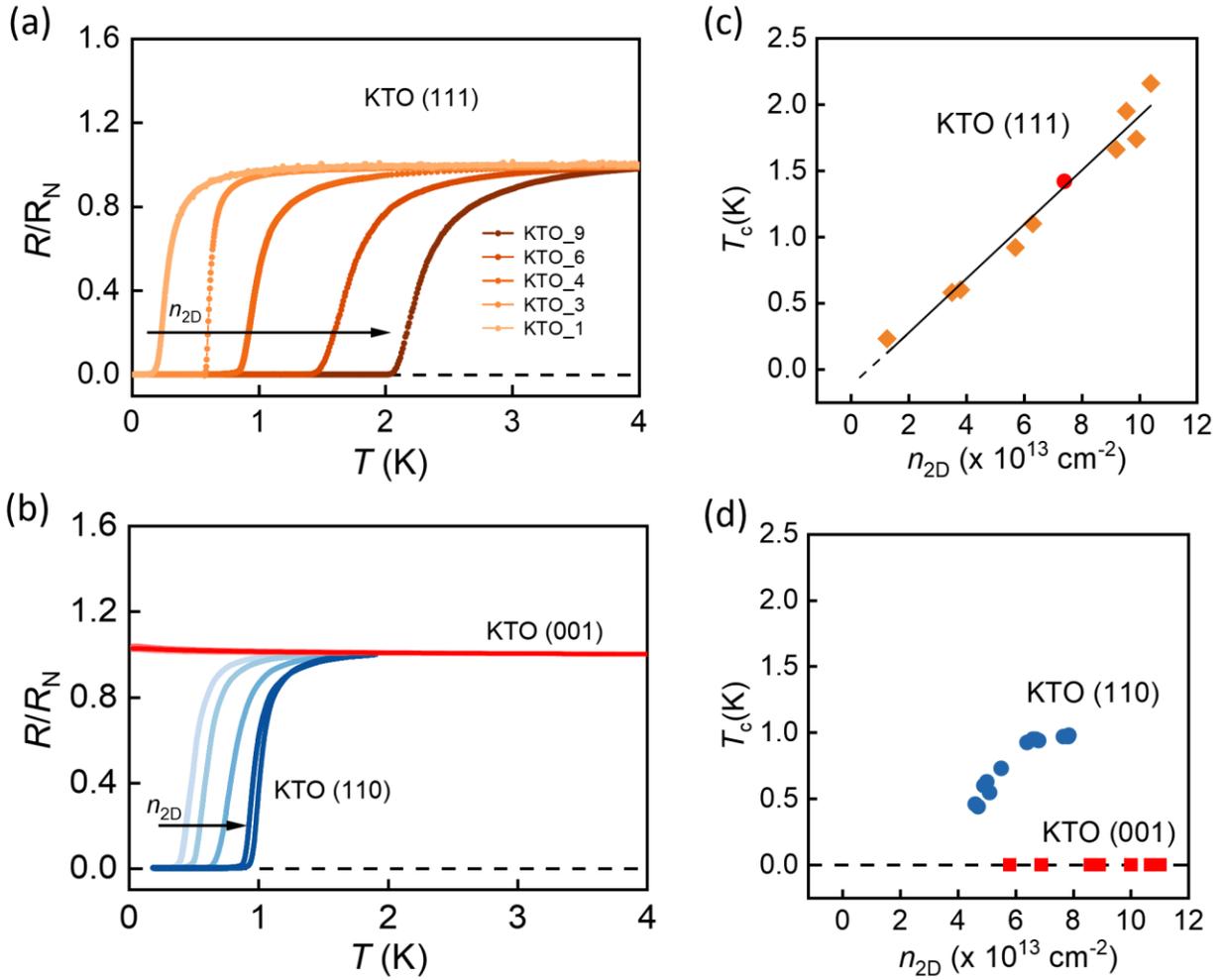

Figure 1. $T_c$ tuned by $n_{2D}$ through chemical doping. (a), (b) The ratio of the resistance $R_s/R_N$ as a function of temperature measured for KTO (111) samples (a) and KTO (001), (110) samples (b), respectively, with varying $n_{2D}$. The direction of the arrow indicates the increase of $n_{2D}$. (c), (d) $n_{2D}$ dependence of $T_c$ for KTO (111) samples shown in (a) and KTO (001), (110) samples shown in (b), respectively. The red circle data point in (c) is taken from Ref. [55].

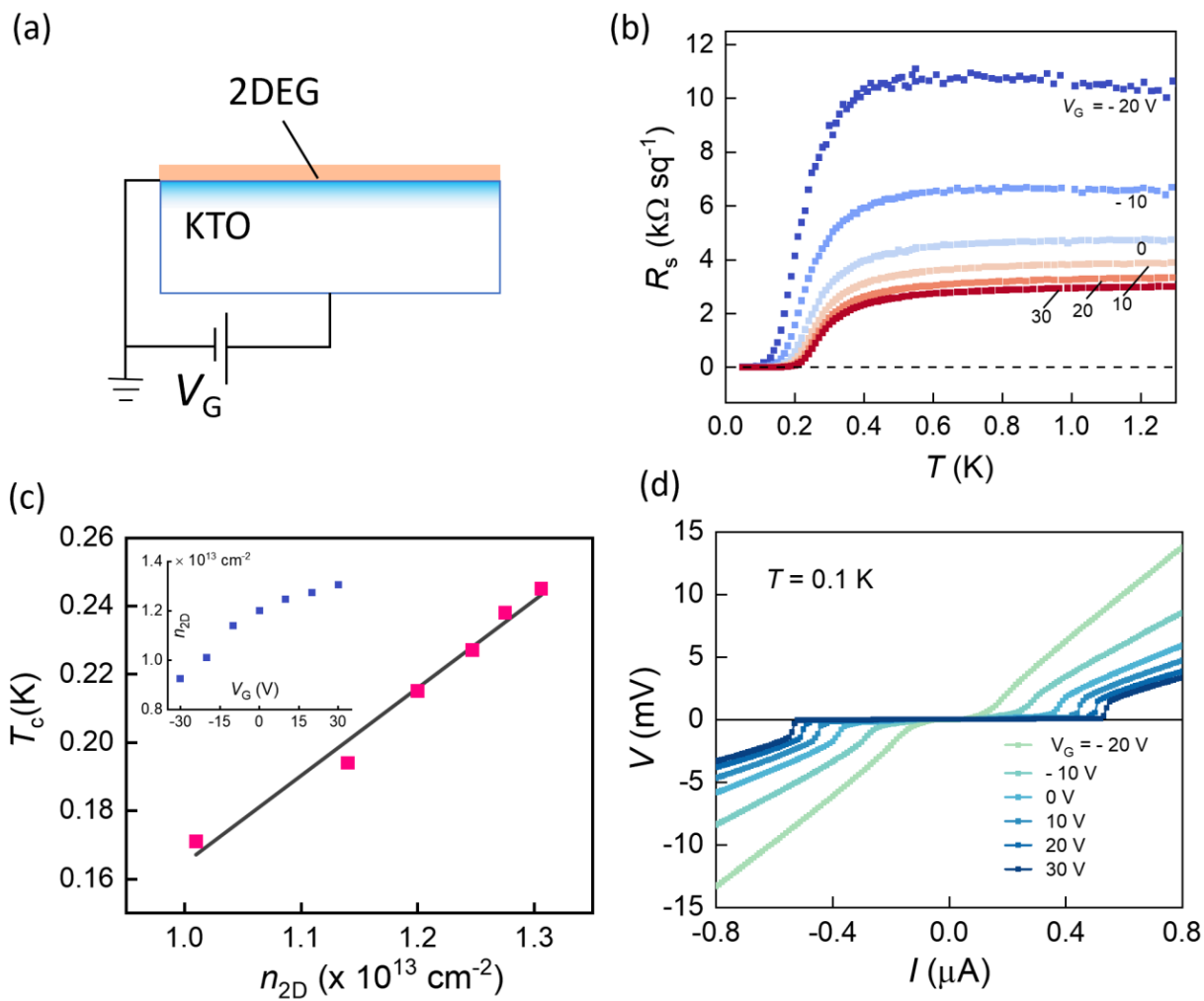

Figure 2. Electrostatic gating on low $n_{2D}$ sample. (a) Schematic of the back gate measurement geometry. (b) $R_s$ versus $T$ measurements on KTO_1 for different $V_G$. (c) $T_c$ versus $n_{2D}$ controlled by electrostatic gating. Inset shows $n_{2D}$ as a function of $V_G$. (d) $V$-$I$ measurement for different $V_G$ at $T = 0.1$ K.

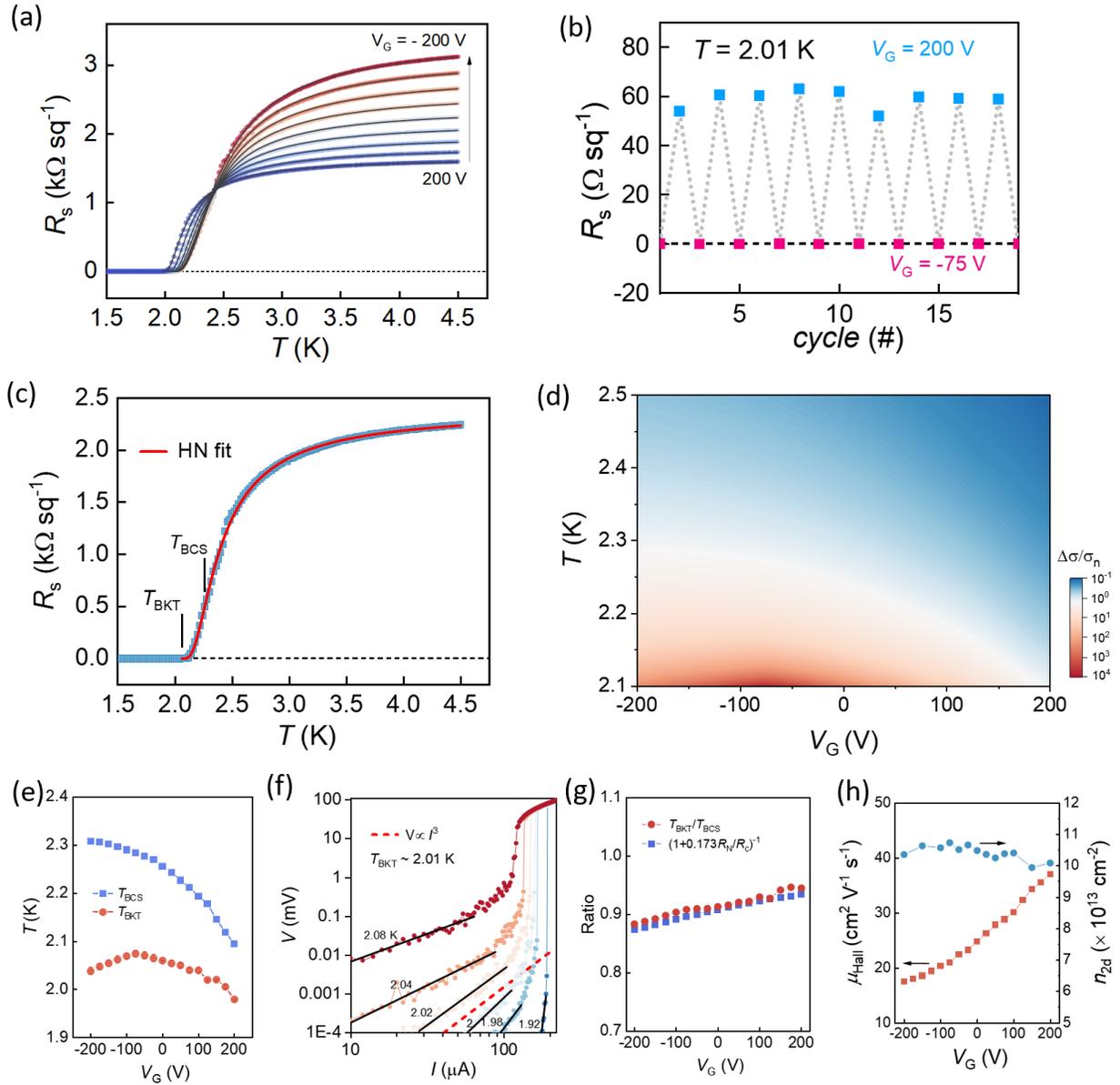

Figure 3. Electrostatic gating for a high $n_{2D}$ sample. (a) Temperature dependence of $R_s$ measured with different $V_G$. The increment in $V_G$ shown in the plot is 50 V. (b) Reversible switching of the superconductivity by $V_G$ at T = 2.01 K. c Halperin-Nelson fit (solid line) to the $R_s(T)$ data. (d) Enhanced conductivity $\Delta\sigma/\sigma_n$ from superconducting fluctuations as a function of both $V_G$ (x-axis) and temperature $T$ (y-axis). (e) Mean-field (blue squares) and BKT (red dots) transition temperatures $T_{BCS}$ and $T_{BKT}$ as a function of $V_G$. (f) Log-Log plot of the $V – I$ measurement; the red dashed line is $V \propto I^3$, indicating $T_{BKT} \sim 2.01$ K. (g) Comparison of the ratio $T_{BKT}/T_{BCS}$ with theory. (h) Hall mobility (left axis) and carrier density (right axis) as a function of $V_G$.

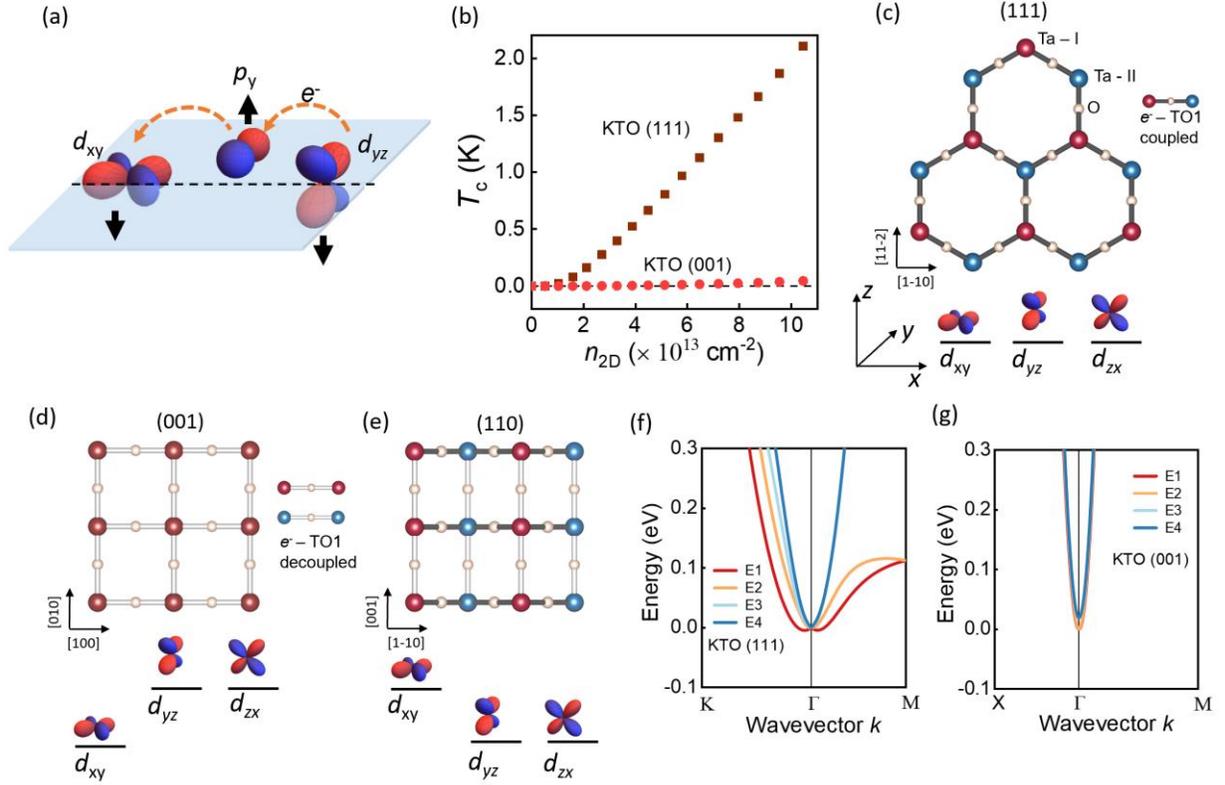

Figure 4. Pairing via transverse optical phonons. (a) Illustration of inter-orbital hopping of electrons between adjacent Ta sites via the O $p_y$ orbital that becomes allowed by inversion symmetry breaking displacements (shown by the black arrows) of O and Ta atoms due to a TO1 phonon. (b) Calculated $T_c$ versus $n_{2D}$ from TO1 phonon pairing for KTO (111) and (001) interfaces, the latter assuming $\lambda_{001}=\lambda_{111}/2$. (c) – (e) Lattice structures formed by the first two layers of Ta and O atoms along the [111], [001] and [110] axes of KTO, respectively. Ta atoms in the upper and lower layers are shown in red and blue, respectively. The dark and light color of Ta-O-Ta bonds indicates the presence or absence of electron-TO1 phonon coupling, respectively. The degeneracy of the $d_{xy}$, $d_{yz}$ and $d_{zx}$ orbitals in energy for each of these KTO surfaces is indicated below each panel by their vertical positions, the degeneracy at the lowest energy (3,1,2) correlating with $T_c$. (f) Rashba-like splitting of bands for a KTO (111) bilayer due to displacements of Ta and O atoms perpendicular to the surface. (g) A similar displacement of atoms for the (001) surface does not produce a noticeable Rashba-like splitting.

**Supplemental figures**

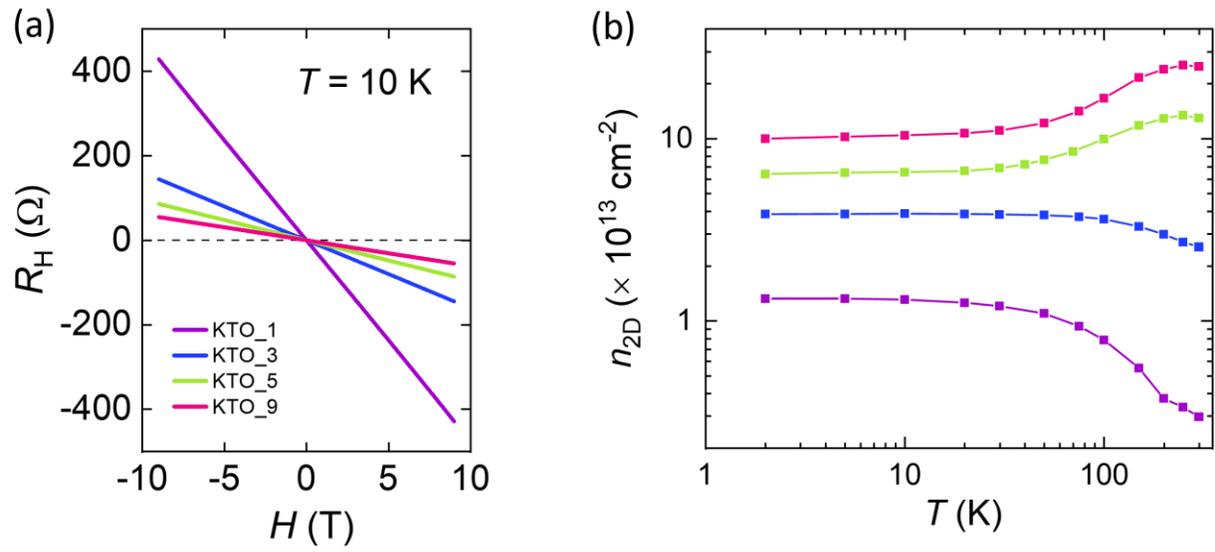

Figure S1. Evolution of donor states for sample with different doping level. (a) Hall resistance as a function of magnetic field for four EuO/KTO (111) samples. (b) Temperature dependence of $n_{2D}$ in samples with different doping levels.

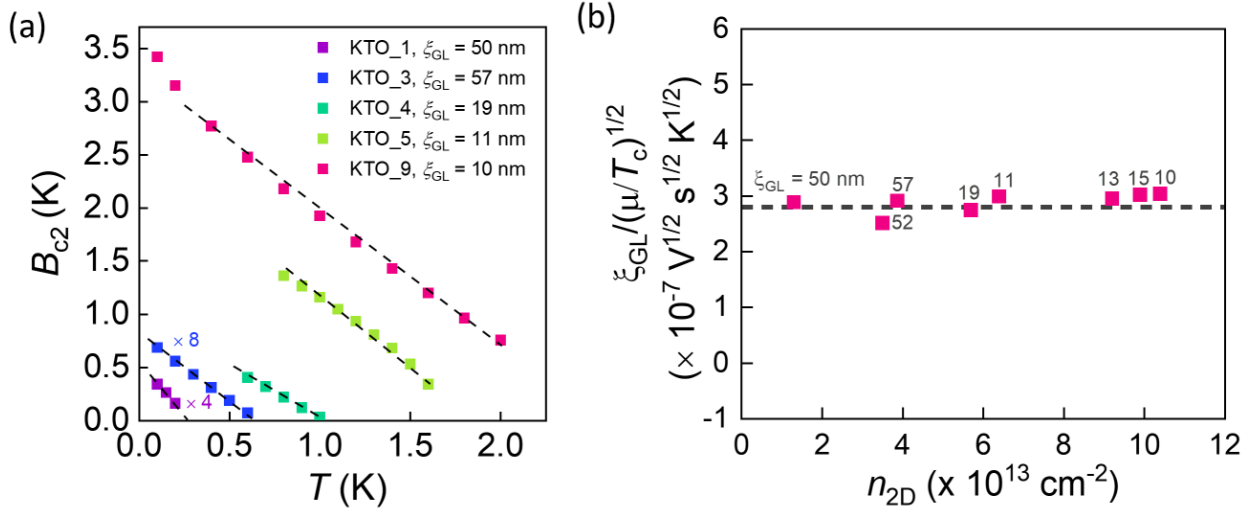

Figure S2. Coherence length measurements. (a) Temperature dependence of the upper critical field measured for EuO/KTO(111) samples with different $n_{2D}$. (b) Scaling of the coherence length $\xi_{GL}$ obtained for different samples. The value of $\xi_{GL}$ is indicated by the number with units of nanometers.

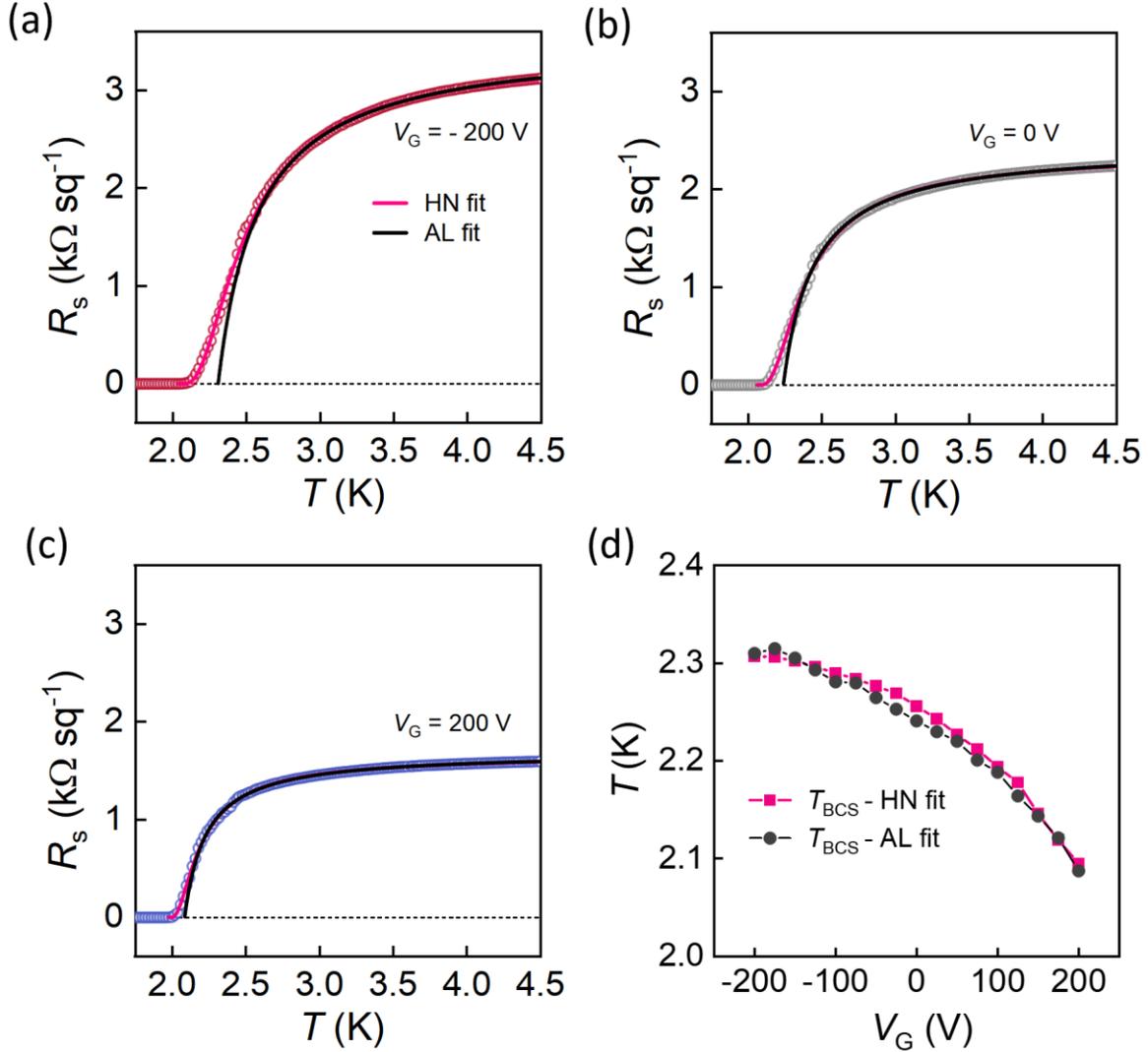

Figure S3. Comparison between the HN and AL fit. (a) – (c) Red and black solid lines are the HN and AL fits to the data for three different $V_G$, respectively. At higher temperatures the two fitted curves are identical to each other. (d) The $T_{BCS}$ obtained from the HN and AL fits for different $V_G$ are shown in red and black, respectively. $T_{BCS}$ obtained from both methods show about the same $V_G$ dependence. The AL fit uses a formula $R_s = 1/((1/R_N) + (1/R_c)T_{BCS}/(T - T_{BCS}))$, where $R_N$ and $R_c$ are fitting parameters. $R_N$ is determined by the normal state resistance of the sample. We note that $R_c$ should be equal to $16\hbar/e^2$ according to the AL formulation. However, we found that the obtained $R_c$ from fitting to the data is less than $16\hbar/e^2$ by about a factor of 2. The reasons for this discrepancy are not known at this time, though processes for conductance enhancement beyond AL (e.g. Maki-Thompson, density of states effects) may play a role.

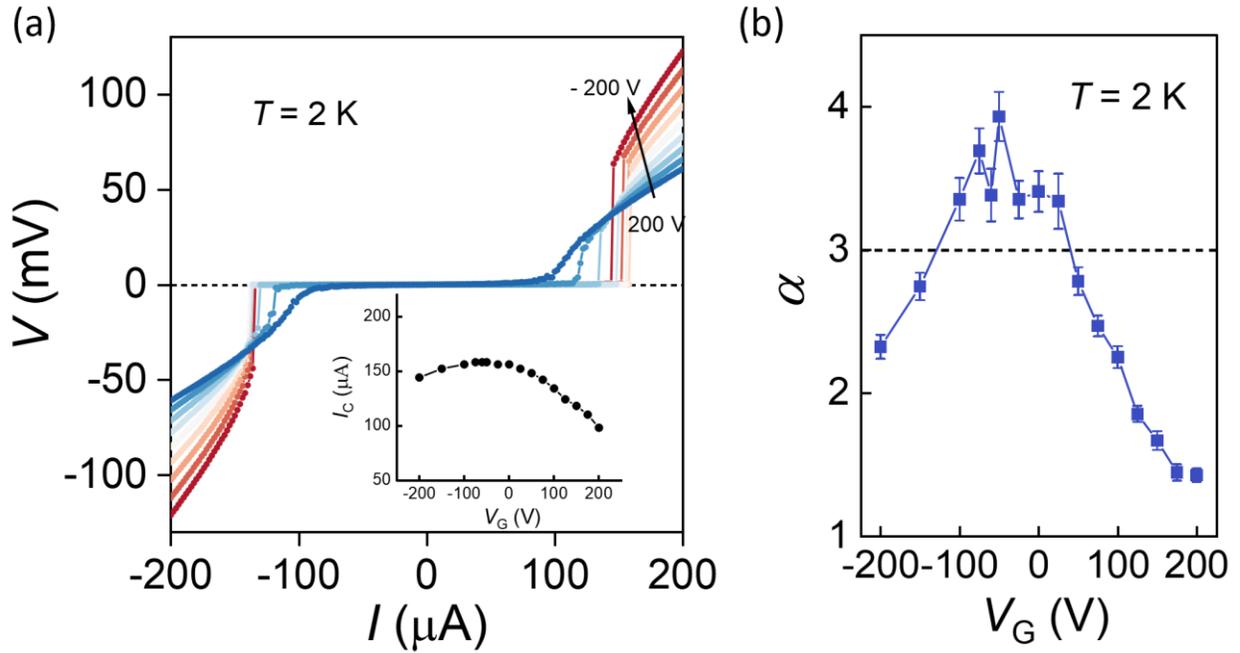

Figure S4. Critical current measurement for different $V_G$. (a) V-I measurements on KTO_9 at $T = 2$ K under different $V_G$. Inset shows the critical current $I_c$ as a function of $V_G$. (b) Exponent $\alpha$ in $V \propto I^\alpha$ as a function of $V_G$. These $V - I$ measurements confirm that the $T_{BKT}$ is tuned by $V_G$. As shown in (b), the exponent $\alpha$ in $V \propto I^\alpha$ obtained in these $V - I$ measurements evolves as a function of $V_G$, showing a local maximum on the negative side of $V_G$. The horizontal dashed line is $\alpha = 3$, which indicates a BKT transition. A value of $\alpha$ that is greater than 3 suggests that the corresponding $T_{BKT} > 2$ K, while $\alpha < 3$ means $T_{BKT} < 2$ K. These $V - I$ measurement results obtained at a fixed temperature for BKT transition are qualitatively the same as those from the HN fit.

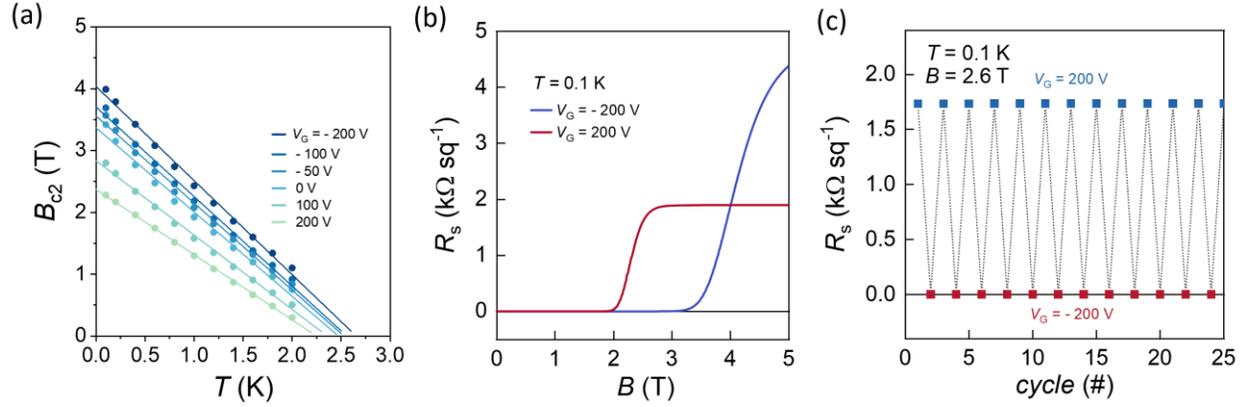

Figure S5. Tuning of critical field and SC on KTO_9 by $V_G$. (a) Temperature dependence of the upper critical field measured at different $V_G$. (b) $R_s$ versus $B$ for two different $V_G$ at $T = 0.1$ K. (c) Switching the SC by $V_G$ at $B = 2.6$ T. Here the critical field at 0 K increases by about a factor of two as seen in (a), which is mainly due to the decrease in mobility from about 37 cm$^2$ V$^{-1}$ m$^{-1}$ to 17 cm$^2$ V$^{-1}$ m$^{-1}$ as $V_G$ changes from 200 V to – 200 V. This is in agreement with the scaling relation $\xi_{GL} \propto (\mu/T_c)^{1/2}$ discussed in the supplemental materials, since $\xi_{GL} \propto B_{c2}^{1/2}$ and the variation in $T_c$ is only ~ 10%. Thus, the tunability of SC shown in (c) is a result of the strong dependence of the critical field on mobility.

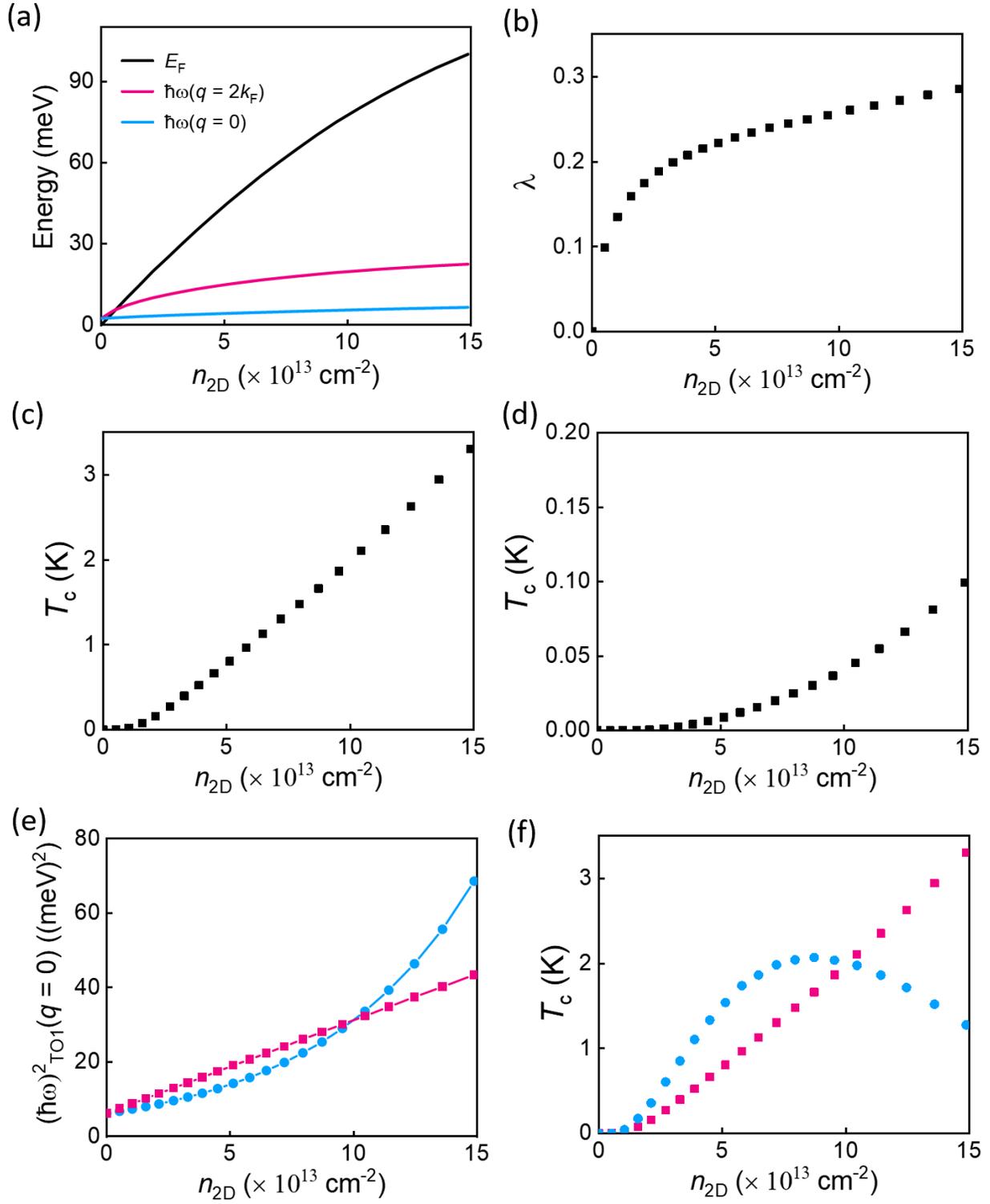

Figure S6. Calculation of TO1 mode energy, BCS coupling constant and $T_c$. (a) Variation versus carrier density for the Fermi energy and TO1 mode energies. (b) BCS coupling constant as a function of $n_{2D}$. (c) Calculation of $T_c$ versus $n_{2D}$ with the prefactor of the coupling constant $c$ set

to match the experimental $T_c$ value of 2 K at $n_{2D} = 10^{14}$ cm$^{-2}$. (d) Calculation of $T_c$ versus $n_{2D}$ with the coupling constant set to half of that in (c). (e) Contrasting dependence of the TO1 mode energy at $q = 0$ on $n_{2D}$ from either a linear assumption (red) or that derived from a triangular potential approximation (blue). (f) Differing dependence of $T_c$ on $n_{2D}$ from (e). The red points in (f) are equivalent to those plotted in (c).

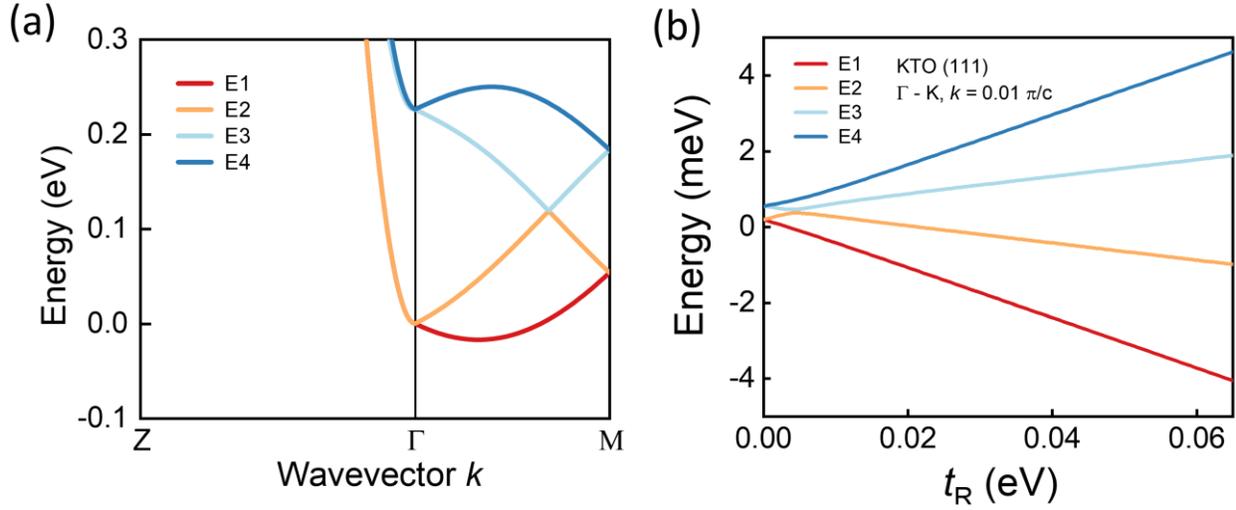

Figure S7. Calculation of the energy dispersion and splitting. (a) Dispersion along symmetry lines of the surface Brillouin zone for a (110) orientation with $t_R = 32.5$ meV. (b) Variation of splitting of the lowest lying quartet versus $t_R$ along $\Gamma$-K for (111), with $c = (2/3)^{1/2}a$. For (110), Z is along [001] and M along [1-10]. Results analogous to (a) for (111) and (001) are shown in Fig. 4f and 4g of the main text, respectively. $t_R$ is the inversion breaking term entering into the tight binding Hamiltonian.

|     | $t_{pd}$ | $\Delta_{pd}$ | $t_{pd}^2/\Delta_{pd}$ | $\eta$ | $\omega_{TO1}$ | $d$ | $g_{TO}/k_F a$ | $t_R$ |
|-----|----------|---------------|------------------------|--------|----------------|-----|----------------|-------|
| STO | 1.136    | 4.018         | 0.321                  | 1.003  | 0.0010         | 0.147 | 0.0968       | 0.0484 |
| KTO | 1.395    | 4.450         | 0.437                  | 0.265  | 0.0025         | 0.060 | 0.0648       | 0.0324 |

Table S1. Calculation of parameters associated with TO1 phonon pairing. $t_{pd}$ is the $pd\pi$ hopping, $\Delta_{pd}$ the energy splitting of the Ta $d$ and O $p$ orbitals, $\eta$ is the ion mass ratio $3M_O/M_{Ta}$, $\omega_{TO1}$ is the bulk TO1 mode energy at $q = 0$, $d$ is the zero-point displacement of the O ions due to the TO1 mode, $g_{TO}$ is the electron-TO1 phonon vertex, and $t_R$ is the inversion breaking term entering into the tight binding Hamiltonian. Here, $k_F$ is the Fermi wavevector and $a$ the bulk lattice constant. Units are eV and Angstroms.

| surface | direction | $\delta/a$ | $\delta_A/a$ | orbitals | $T_c$ |
|---|---|---|---|---|---|
| (111) | Γ-K | 1.712 | $3^{1/2}$ | xy, xz, yz | ~ 2 K |
|  | Γ-M | 1.716 | $3^{1/2}$ |  |  |
| (001) | Γ-X | 0.453 | $2\xi/\varepsilon$ | xy | ~ 0 K |
| (110) | Γ-Z | 0.000 | 0 | xz, yz | ~ 1 K |
|  | Γ-M | 1.414 | $2^{1/2}$ |  |  |

Table S2. Comparison of inversion splitting for the three crystallographic orientations. $\delta$ for $k$ near Γ, with the inversion dependence of the lowest band energy going as $-\delta t_R k$, with $\delta_A$ a simple analytic estimate in the limit that $k$ goes to 0. "orbitals" denote those $t_{2g}$ orbitals present in the lowest energy bands. $T_c$ is the maximally observed $T_c$ for a given interface orientation. Here, a is the bulk lattice constant, $\xi$ the spin-orbit coupling, and $\varepsilon$ the energy splitting between the xy and xz/yz states.

**Acknowledgements**

We thank Ivar Martin, Peter Littlewood, Alex Edelman, Arun Paramekanti and James Rondinelli for discussions. All research presented here is supported by the Materials Science and Engineering Division, Office of Basic Energy Sciences, U.S. Department of Energy. The use of facilities at the Center for Nanoscale Materials was supported by the U.S. DOE, BES under Contract No. DE-AC02-06CH11357.

**Author contributions**

Synthesis of samples was carried out by C.L. with assistance from D.H., J.P., H.Z. and A.B.. Measurements in PPMS and with the adiabatic demagnetization option were carried out by C.L. with assistance from B.F. and J.P.. Measurements in the dilution fridge at the Center for Nanoscale Materials were carried out by C.L. with assistance from X.Z. and D.J.. Analysis of the data was carried out by C.L. with assistance from M.N. and A.B.. All theoretical work presented here was carried out by M.N.. A.B. supervised all experimental work and analysis. The paper was written by C.L, M.N. and A.B.. All authors contributed to discussions regarding the paper.